# Minimization of Eddy Current Artifacts in Sequences with Periodic Dynamics

Sebastian Flassbeck*[1,2] | Jakob Assländer[1,2]

[1]Center for Biomedical Imaging, Dept. of Radiology, New York University Grossman School of Medicine, NY, USA

[2]Center for Advanced Imaging Innovation and Research (CAI2R), Dept. of Radiology, New York University Grossman School of Medicine, NY, USA

**Correspondence**
*Sebastian Flassbeck, Center for Biomedical Imaging, Department of Radiology, New York University Grossman School of Medicine, 650 1st Avenue, New York, NY 10016, USA.
Email: sebastian.flassbeck@nyulangone.org

**Abstract**

**Purpose:** To minimize eddy current artifacts in periodic pulse sequences with balanced gradient moments as, e.g., used for quantitative MRI.

**Theory and Methods:** Eddy current artifacts in balanced sequences result from large jumps in k-space. In quantitative MRI, one often samples some spin dynamics repeatedly while acquiring different parts of k-space. We swap individual k-space lines between different repetitions in order to minimize jumps in temporal succession without changing the overall trajectory. This reordering can be formulated as a traveling salesman problem and we tackle the discrete optimization with a simulated annealing algorithm.

**Results:** Compared to the default ordering, we observe a substantial reduction of artifacts in the reconstructed images and the derived quantitative parameter maps. Comparing two variants of our algorithm, one that resembles the *pairing* approach originally proposed by Bieri et al., and one that minimizes all k-space jumps equally, we observe slightly lower artifact levels in the latter.

**Conclusion**: The proposed reordering scheme effectively reduces eddy current artifacts in sequences with balanced gradient moments. In contrast to previous approaches, we capitalize on the periodicity of the sampled signal dynamics, enabling both efficient k-space sampling and minimizing artifacts caused by eddy currents.

## 1 | INTRODUCTION

Balanced steady-state free precession (bSSFP) sequences,[1] i.e., sequences with balanced gradient moments or simply "balanced sequences," are known for their unparalleled SNR efficiency. This beneficial property is commonly utilized for cine cardiac imaging and can be translated to transient- or hybrid-state sequences, for example, for quantitative MRI.[2–6] In dynamic imaging, it is often unfeasible to Nyquist-sample k-space for each time frame due to constraints on the measurement time. One can utilize correlations between different time frames,[3,7,8] but this necessitates the acquisition of complementary k-space lines across different time frames.

Heuristically, it is often assumed that consecutive time points contain the most correlated signal, which motivated the development of the golden angle scheme.[3,9] However, this assumption is inaccurate in periodic dynamics, where, by definition, identical signals are measured repeatedly. To illustrate the inefficiency of this approach, we can assume a repeated sampling of some dynamics with a periodicity of $N_t$.

In this case, the angular increment within each time frame is $N_t \cdot \phi_g$, where $\phi_g$ denotes the golden angle. Assuming, e.g., $N_t = 233$ results in a phase increment of 0.3 degree, which is clearly a sub-optimal sampling scheme. One approach to guarantee an efficient sampling is to exploit the periodicity and reorder the golden angle series such that the golden angle increments occur first between different repetitions of the same time frame and, in an outer loop, between neighboring time frames (Fig. 1).

The original golden angle approach entails large jumps in the trajectory between adjacent $T_R$s, which induce strong temporal variations of eddy currents, causing phase fluctuations, and ultimately result in image artifacts when paired with balanced sequences.[10-14] The tiny golden angle scheme[10] overcomes this problem, but this approach is ineffective for the above-described reordered sampling scheme as the increment in temporal succession is $N_c$ times the angle increment where $N_c$ denotes the total number of periodic dynamics acquired, termed *cycles*.

Fortunately, the repeated sampling of the same spin dynamics disentangles the temporal succession of the k-space trajectory—which is relevant for eddy-current artifacts—and the golden angle index, which is relevant for efficient encoding of the k-t-space. In this work, we utilize the interchangeability of the k-space acquisitions between repetitions of the same time frame and formulate an optimization problem that minimizes eddy-current artifacts without changing the overall k-space trajectory and, therefore, without compromising the encoding efficiency (Fig. 1). Further, we demonstrate that this concept generalizes to many k-space trajectories beyond a reordered golden-angle scheme.

## 2 | THEORY

Gradient pulses generate eddy currents on the conducting surfaces of the MRI hardware, which, in turn, cause unwanted temporally and spatially variable magnetic fields. Bieri et al.[11] showed that the steady state in bSSFP sequences is not disrupted by eddy-current-induced phase variations evoked by gradient pulses that are repeated every $T_R$. This condition is approximately fulfilled by k-space trajectories with linear phase encoding. In contrast, gradient pulses that substantially change in amplitude or direction between $T_R$s do disrupt the steady state by imparting an additional phase to the magnetization. This phase causes oscillations of the magnetization around the steady-state cone, which results in image artifacts.[15,16]

bSSFP sequences cycle on-resonant magnetization around the z-axis every $T_R$. This cyclic behavior of the magnetization is utilized by the pairing approach for phase encoding (PE) gradients proposed by Bieri et al.[11]: two adjacent k-space lines are acquired in consecutive $T_R$s before an arbitrary jump in k-space is allowed. They showed that the eddy current induced phase accumulation over two consecutive $T_R$s with similar k-space trajectories mostly cancel each other out. This concept provides some flexibility for the design of k-space sampling schemes but entails a constraint on the k-space jump between every other $T_R$, which potentially limits the encoding efficiency.

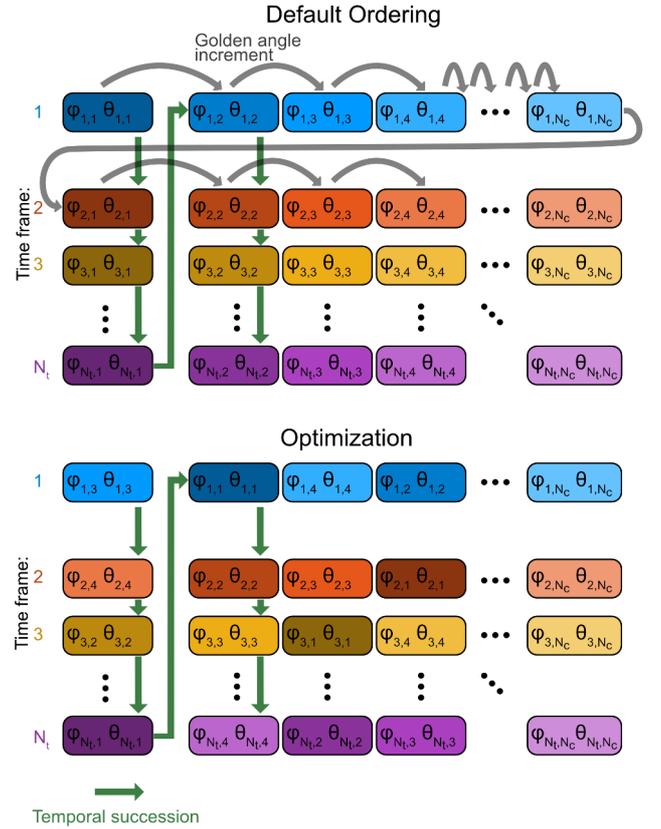

**FIGURE 1** Sketch of the golden angle ordering (top) and the proposed optimized re-ordering (bottom). Readouts colored in the same hue belong to a single time frame. In the default ordering, the color luminosity encodes the different cycles. After the optimization, however, readouts are reordered within each time frame. Note that the golden angle series runs over the cycles first before going to the next time frame (gray arrows), while the time series is sampled first in the experiment before acquiring the next cycle (green arrows). The optimization changes the acquisition order by permuting along the cycles-direction (same hue, different luminosity), ensuring that the overall k-space information for every time frame remains unchanged.

## 3 | METHODS

### 3.1 | Pulse sequence

To demonstrate the proposed reordering scheme, we used the hybrid-state free precession (HSFP) sequence described in Ref. [6], which employs non-selective rectangular RF-pulses

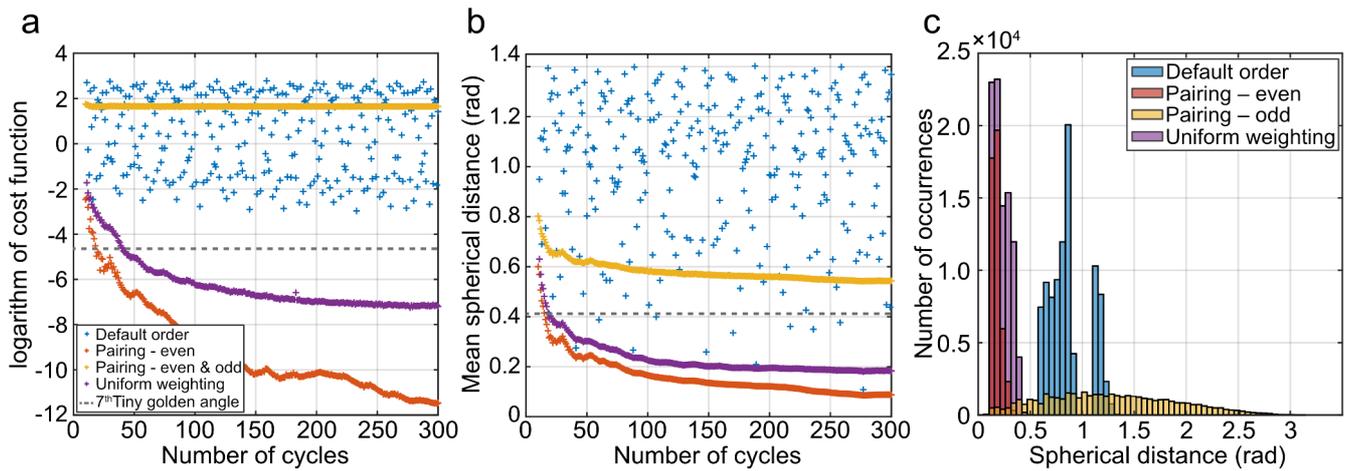

**FIGURE 2 a**: Cost function before and after the simulated annealing optimization, which uses $p = 6$ (cf. Eq. (3)). With a uniform weighting every angle increment is considered equally, while in the pairing approach, only every other angle increment contributes to the cost function (red line). The yellow line evaluates the pairing optimization with uniform weighting, i.e., considering even and odd jumps. **b**: Comparison of the mean spherical distance of the different k-space trajectories. For comparison, we added the cost function and spherical distance between adjacent spokes of the 2D $7^{th}$ tiny golden angle (dashed lines), which is known for its benign eddy current properties.[10] **c**: Histogram of the spherical distance for $N_c = 94$, as used for the experiments. The histogram is split between even and odd increments only for the pairing case, as the distributions for the default order and the uniform optimization exhibit virtually no difference between even and odd pairs.

and performs k-space encoding with a 3D koosh-ball trajectory and with a nominal isotropic resolution of 0.8 mm. The complete list of sequence parameters and a plot of the flip angle pattern are shown in the Supporting Information. As typical for hybrid-state sequences, the gradient moments in each $T_R$ are fully balanced and the magnetization is driven into a dynamic equilibrium by continuously changing the flip angle. This dynamic equilibrium with a cycle duration of 3.8 seconds was repeated 94 times ($N_c = 94$) and different parts of k-space were sampled in each cycle. The overall measurement time was 6 minutes.

The flip angle pattern used in this work was optimized for the quantification of $T_1$ and $T_2$. However, we reduced the $T_R$ to 3.9 ms for the present work and linearly interpolated the flip angle pattern to ensure a largely unchanged temporal evolution of the magnetization according to the hybrid-state theory.

## 3.2 | k-Space trajectories

Here, we demonstrate the proposed approach to reduce eddy current artifacts for an HSFP sequence with a radial k-space readout where the readout direction was incremented by a reordered 2D golden angle scheme,[17] which is described above and is illustrated in Fig. 1.

For a sequence with $N_t$ time points and $N_c$ cycles, the polar angle $\theta_{t,c}$ and the azimuth $\varphi_{t,c}$ of the $t^{th}$ time frame and the $c^{th}$ cycle are given by

$$\theta_{t,c} = \cos^{-1}(\text{mod}((N_c \cdot t + c) \cdot \phi_1, 1)) \quad (1)$$

$$\varphi_{t,c} = (N_c \cdot t + c) \cdot 2\pi \cdot \phi_2. \quad (2)$$

Here, $\phi_1 \approx 0.4656$ and $\phi_2 \approx 0.6823$ are the 2D-golden mean ratios.[17] If we denote the readout directions in a matrix of size $N_t \times N_c$, the temporal readout succession is the *column major order*, i.e. we first acquire the first column from the top to the bottom, followed by the second column and so forth. In a perfect MRI system, permuting the elements in each row, i.e. changing the order of the index $c$ of $\{\theta_{t,c}, \varphi_{t,c}\}$, would have no effect as, overall, the same information is measured. However, with a real system, this order does impact the eddy currents and their effect on the spin dynamics.[11,13] We utilize this property to formulate an optimization problem in search of minimal eddy current artifacts without altering the overall k-space trajectory.

## 3.3 | Simulated annealing

Given a predefined set of k-space lines or spokes in a dynamic but periodic imaging experiment, the temporal order in which these spokes are acquired can be viewed as a discrete optimization problem. The aim of this optimization is then to minimize a distance metric between the starting points of temporally adjacent spokes. As we are measuring all spokes in the predefined set, just in an optimized order, we can view this optimization as a version of the well-known traveling salesman problem, which is NP-hard but approximate solutions can be found, for example, with simulated annealing (SA).

We limit the description of simulated annealing algorithms to our concrete problem and we refer to the literature, e.g., the textbook by van Laarhoven et al.[18] for further details. In SA algorithms, one introduces a random perturbation of the current state and compares an objective function before and after this perturbation. In our optimization problem, the perturbation is swapping the order of acquisition of two spokes from cycle $c$ and cycle $\tilde{c}$ for a given time frame $t$. The objective function $\mathcal{F}$ is given by:

$$\mathcal{F} = \sum_{c=1}^{N_c} \sum_{t=1}^{\frac{N_t}{2}-1} w_1 \left\| \vec{k}_{2t-1,c} - \vec{k}_{2t,c} \right\|_2^p + w_2 \left\| \vec{k}_{2t,c} - \vec{k}_{2t+1,c} \right\|_2^p \quad (3)$$

with

$$\vec{k}_{t,c} = \begin{pmatrix} \cos\varphi_{t,c}\,\sin\theta_{t,c} \\ \sin\varphi_{t,c}\,\sin\theta_{t,c} \\ \cos\theta_{t,c} \end{pmatrix}. \quad (4)$$

The exponent $p$ allows for a trade-off between reducing the total distance traversed by the trajectory—which would be achieved by $p = 1$—and focusing on the largest jump—which would be achieved by $p = +\infty$. An analysis (see Supporting Information) of the exponent identified $p = 5$ as optimal with only minor differences in the range 4–6 and we heuristically chose $p = 6$. In our implementation, the SA algorithm is performed for a fixed number of iterations ($N_{\text{iterations}} = 10^9$), which takes approximately 5 minutes for our use case on a 2015 Intel i7 processor (single core). The simulated annealing algorithm is illustrated in Alg. 1.

**Algorithm 1** Pseudo-code of the simulated annealing algorithm.

---
$I \leftarrow 0$
**while** $I < N_{\text{iterations}}$ **do**
    $\mathcal{F} \leftarrow$ Eq. (3)
    $t \leftarrow$ random index $\in \{1, \ldots, N_t\}$
    $c \leftarrow$ random index $\in \{1, \ldots, N_c\}$
    $\tilde{c} \leftarrow$ random index $\in \{1, \ldots, N_c\}$
    swap $\vec{k}_{t,c}$ and $\vec{k}_{t,\tilde{c}}$
    $\tilde{\mathcal{F}} \leftarrow$ Eq. (3)
    $u \leftarrow$ random number $\in [0, 1]$
    **if** $\exp(-\frac{\tilde{\mathcal{F}}-\mathcal{F}}{((1-I/N_{\text{iterations}})^{p_t})}) > u$ **then**
        keep the swap of $\vec{k}_{t,c}$ and $\vec{k}_{t,\tilde{c}}$
    **else**
        undo the swap of $\vec{k}_{t,c}$ and $\vec{k}_{t,\tilde{c}}$
    **end if**
    $I \leftarrow I + 1$
**end while**

---

The cost function (Eq. (3)), provides the flexibility to vary the weights $w_{1,2}$. In this article, we explore two edge cases. First, we analyze the setting $w_1 = 0$ and $w_2 = 1$, penalizing only the readout direction changes between every other $T_{\text{R}}$. The resulting temporal order of projections resembles the pairing approach proposed by Bieri et al.[11] while maintaining the same overall k-space trajectory. Effectively, this version of our approach generalizes the pairing approach to a periodic multi-k-space setting and provides an approximate solution to the NP-hard optimization problem. For comparison, we analyze the setting $w_1 = w_2 = 1$, which uniformly penalizes all k-space jumps.

We note that it is computationally inefficient to calculate $\mathcal{F}$ (Eq. (3)) before and after a swap since most terms in those sums remain unchanged. For this reason, we compute only the eight terms affected by the swap (cf. the code referenced in Appendix A).

### 3.4 | Phantom and in vivo experiments

We validated the proposed approach by comparing phantom and in vivo scans with the default ordering to optimized ordering schemes with the aforementioned pairing and uniform weighting. We scanned the NIST system phantom (NIST, Boulder, CO, USA)[19] and three healthy volunteers, after getting informed consent in agreement with our IRB. These scans were performed on a 3T PRISMA (Siemens Healthineers, Erlangen, Germany) using a 32-channel head coil for the in vivo scan and a 20-channel head coil for the phantom experiments. Further, we scanned the NIST phantom on a 1.5T Sola system (Siemens Healthineers, Erlangen, Germany), where the relaxation times were compared to the vendor-provided reference values (Supporting Figs. S8 & S9).

### 3.5 | Reconstruction and data analysis

Heuristically, we find that the hybrid-state signal is low rank. E.g., for the current use case, 99.3% of the signal energy from the dictionary described below are represented by only 5 basis functions. We make use of this low-rank data structure and reconstruct images directly in this 5-dimensional sub-space. The basis functions that span this space were calculated by a singular value decomposition of the dictionary, which is also used for parameter estimation. We reconstructed 5 *coefficient images* that correspond to the 5 basis functions by solving the *low-rank inverse problem* described in Refs. 7, 8 with an additional locally-low rank penalty[20]:

$$\min_{\tilde{\mathbf{x}}} \|\mathbf{E} \cdot \mathbf{U} \cdot \tilde{\mathbf{x}} - \mathbf{s}\|_2^2 + \lambda \sum_r \|R(\tilde{\mathbf{x}}_r)\|_* . \quad (5)$$

Here, $\mathbf{E}$ is the encoding matrix, $\mathbf{U}$ the subspace projection matrix composed of the basis functions, $\mathbf{s}$ the measured signal, and $\tilde{\mathbf{x}}$ the coefficient images. The second term is a locally low-rank regularizer, where $\lambda$ is the regularization parameter, $R$ is an operator that extracts blocks from the coefficient images and builds a matrix where each column spans the coefficients and each row of the voxels in a particular block $r$. $\| \cdot \|_*$ denotes the nuclear norm. We used the Berkeley Advanced Reconstruction Toolbox[21] to solve this reconstruction problem. All measurements were reconstructed using the same reconstruction pipeline with a heuristically chosen $\lambda = 10^{-5}$.

We performed dictionary matching to estimate the parameter maps. The dictionary was calculated with the Bloch model for relaxation times in the ranges $T_1 \in [0.1\,\mathrm{s}, 3\,\mathrm{s}]$ and $T_2 \in [0.01\,\mathrm{s}, 2\,\mathrm{s}]$, exponentially spaced with 1% increments. Nominal values for $B_0$ and $B_1$ were assumed and a single on-resonant isochromat was simulated for each pair of relaxation times. We projected the dictionary to the sub-space spanned by $\mathbf{U}$, which allows for a direct matching of the coefficient images.[22]

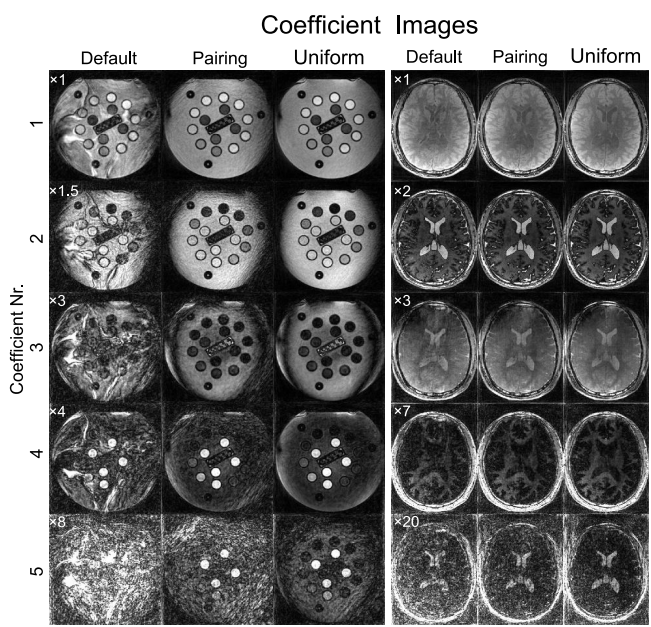

**FIGURE 3** The magnitude of the coefficient images for the phantom and in vivo scans. Each column corresponds to a separate scan with the respective ordering scheme. The rows show the 5 coefficients reconstructed from each scan[8]. The multiplicative factors in the upper left corner show the scaling of the images relative to the first coefficient.

## 4 | RESULTS

Fig. 2a demonstrates the effectiveness of the simulated annealing algorithm in reducing k-space jumps, in particular for $N_c \gtrsim 20$. For smaller values of $N_c$, the limited options for permutation prevent a more effective optimization. A comparison of the sub-figures a and b, which correspond to the cost function in Eq. (3) with $p = 6$ and the mean spherical distance reveals that the algorithm predominantly reduces large jumps and to a lesser degree the mean distance. As expected, the pairing approach shows smaller values in both metrics if we only consider the distance within each pair. For a more detailed analysis, Fig. 2c depicts histograms of the spherical distances for $N_c = 94$, which corresponds to the setting used in the measurements. The histograms again show the reduction in the k-space jumps and furthermore reveal the difference between pairing and the distribution from the uniformly weighted optimization.

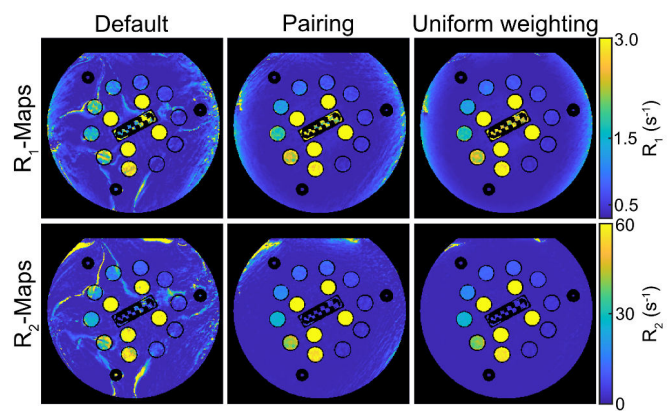

**FIGURE 4** Quantitative relaxation rate maps of the NIST system phantom, estimated from the coefficient images shown in Fig. 3. The black areas masked are based on the thresholding of the coefficient images.

The coefficient images of the phantom scan in Fig. 3 contain severe artifacts when using the default ordering. These artifacts are mitigated well when reordering the k-space spokes for pairing and best with uniform weighting. For the default ordering scheme, the artifacts propagate to the quantitative maps. For the paired reordering, we observe an increased apparent noise level compared to the uniform weighting scheme (Fig. 4).

The coefficient images of the in vivo scan exhibit a similar trend, although the eddy-current artifacts are generally less severe compared to the phantom scans. This difference is likely caused by the long relaxation times in the main water compartment of the latter. Nevertheless, an improvement in the image quality was achieved in vivo with both weightings and, in particular, using the uniform weighting (coefficient #5 in Fig. 3). This improvement translates directly to the parameter maps (Figs. 4 and 5).

All parameter maps show a spatially varying bias, which is likely due to $B_0$ and $B_1$ inhomogeneities. Most prominent is the gradual change of $T_1$ in the main body of the NIST-phantom, but we also observe increased relaxation rates in the prefrontal cortex in vivo. For this reason, we performed a quantitative analysis at 1.5T, where $B_0$ and $B_1^+$ fields are more homogeneous. In this analysis, the mean $T_1$ estimates based on default order, pairing, and the uniform pattern deviated from NMR measurements by 14.9%, 9.1%, and 8.0%, respectively, and the mean $T_2$ estimates by 32%, 21%, and 9.5% (cf. Supporting Figs. S8 & S9).

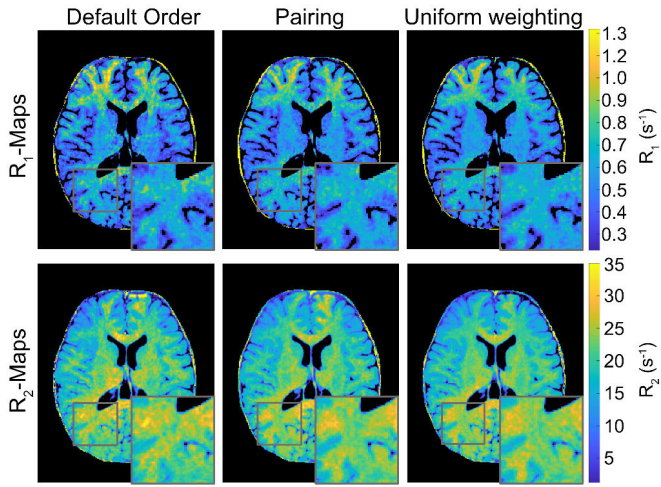

**FIGURE 5** Quantitative maps of the relaxation times in a representative transverse slice of the 3D full brain scan. The magnification highlights diffuse eddy-current artifacts that are mitigated by the pairing flavor of our approach and, to a larger extent, by the uniform scaling flavor of our approach. Two additional volunteer scans can be found in the Supporting Information (Figs. S6 & S7).

We note that in vivo parameter estimates are biased by magnetization transfer. A model that accounts for magnetization transfer is a subject of our ongoing research, as is the correction of $B_0$, $B_1$ inhomogeneities.[23–25] However, these topics are beyond the scope of this paper.

## 5 | DISCUSSION

We formulated the mitigation of eddy-current artifacts as a traveling salesman problem and solved it with simulated annealing. We demonstrated that this approach reduces the artifact level (Figs. 3–5). Unlike previous approaches like tiny golden angles or Bieri's original line-pairing, the proposed approach does not sacrifice encoding efficiency.

An analysis of different flavors of the proposed approach showed that the uniform weighting outperforms the paired weighting. However, this performance gap is significantly smaller compared to the improvement over the default ordering. Since the pairing approach had the overall smallest cost and mean spherical distance (cf. Fig. 2), the improvement obtained with the uniform weighting shows that the assumption that eddy-current-induced phases cancel out over two $T_R$s is not exactly fulfilled. We note that both variants of the proposed approach studied here are edge cases and the formalism allows for any combination of $w_{1,2}$ (Eq. (3)).

Though the artifact level in our example application was relatively insensitive to $w_{1,2}$, smaller $N_c$ values might exhibit more sensitivity and an intermediate weighting might be optimal. The fine-tuning of these settings is likely dependent on the specific magnetization dynamics and the k-space sampling scheme. For this reason, we forgo a more detailed analysis. To demonstrate the generalizability of the proposed algorithm to virtually any k-space trajectory, we study an inversion recovery bSSFP experiment with a Cartesian readout and random phase encoding in the Supporting Information.

The current paper focuses on quantitative MRI, which we believe is the most obvious application for the proposed approach. However, the method might also be beneficial for cine cardiac imaging, where tiny-golden angle schemes are commonly used.[10] As discussed above, a tiny golden angle pattern in temporal succession has suboptimal encoding efficiency which, in this case, depends on $N_t$ and therefore the heart rate. With the here-proposed approach, one could sample k-space with golden angle increments along the axis of different heartbeats and, in an outer loop, along time within each heartbeat. This setup would set the stage for the proposed reordering scheme to minimize eddy currents. However, this requires the sequence to be updated prospectively to deal with the diastolic variability of the cardiac cycle. One would likely need to incorporate a buffer time in which the same or similar k-space spokes are acquired repeatedly until a new heartbeat is detected, which introduces a trade-off between the encoding efficiency in systole and diastole. Such an approach necessitates real-time adjustment of the acquisition, which makes it incompatible with retrospective gating.

## 6 | CONCLUSION

Simulated annealing effectively reduces eddy current artifacts without changing the overall k-space trajectory and, hence, without impairing its encoding efficiency. The proposed framework builds on the repeated sampling of the same dynamics with different k-space encoding and can be combined with different variants of Cartesian and non-Cartesian trajectories.

## ACKNOWLEDGEMENTS

The authors would like to thank Andrew Mao for proofreading the manuscript.

# APPENDIX

## A DATA AVAILABILITY STATEMENT

In order to ensure reproducibility, we provide the following resources:

- An implementation of the simulated annealing algorithm can be found on https://github.com/JakobAsslaender/MRIeddyCurrentOptimization.jl. It is written in the open source language Julia and the package can be installed via Julia's package manager with the command: "]add https://github.com/JakobAsslaender/ MRIeddyCurrentOptimization.jl".

- A documentation of the package with example scripts can be found on: https://jakobasslaender.github.io/MRIeddyCurrentOptimization.jl, where we also linked to a Jupyter notebook that can be launched in *binder*, allowing for an interactive exploration of the algorithm in a browser without any local installations.

All figures in this paper were created with version v0.2.0 (commit hash e7f70c3) of the MRIeddyCurrentOptimization.jl package and match the version's documentation.

## ORCID

*Sebastian Flassbeck*\*  0000-0003-0865-9021
*Jakob Assländer*  0000-0003-2288-038X